\documentclass[nofootinbib]{andromedaone}
\usepackage{amsmath,amssymb}
\usepackage{hyperref}
\usepackage[outdir=./]{epstopdf}
\journal{BSM}
\vol{2023}
\jyear{Egypt}
\pages{Hurghada}

\biboptions{sort&compress}
\received{xx March 2020}
\published{xx March 2020}

\def\be{\begin{equation}}
\def\ee{\end{equation}}
\def\bea{\begin{eqnarray}}
\def\eea{\end{eqnarray}}

\begin{document}

\title{Domain Walls in the $A_4$ flavored NMSSM}

\author{Mohamed Amin Loualidi\auno{1,*}}
\author{and Salah Nasri\auno{1,2,+}}
\address{$^1$Department of physics, United Arab Emirates University, Al-Ain, UAE}
\address{$^2$The Abdus Salam International Centre for Theoretical Physics, Strada Costiera 11, I-34014, Trieste, Italy}
\address{Email addresses: \textbf{*} ma.loualidi@uaeu.ac.ae, \textbf{+} snasri@uaeu.ac.ae, salah.nasri@cern.ch}

\begin{abstract}
In this work, we study the phenomenology of neutrinos and the formation of cosmic domain walls in the NMSSM extended by an $A_4 \times Z_3$ flavor symmetry. Neutrino masses result from the type I seesaw mechanism using only two flavon fields and the NMSSM singlet $\mathcal{S}$ while their mixing is of Trimaximal mixing form. We perform our phenomenological study in the normal mass hierarchy where we find that observables like $m_{\beta\beta}$, $m_{\beta}$, and $\sum m_i$ can be tested by future experiments. Due to the difference between the $A_4$ subgroups that undergo spontaneous breaking in both the charged lepton and neutrino sectors, the resulting domain walls in each sector exhibit distinct structures. We delve into the details of the breaking patterns within these two sectors, and we introduce a nuanced geometric representation for them. To tackle the domain wall problem, we explore a well-established method involving the explicit breaking of the flavor symmetry. This is achieved through the introduction of Planck-suppressed operators induced by supergravity.
\end{abstract}

\maketitle

\section{Introduction}
The interplay between spontaneous symmetry breaking (SSB), discrete symmetries, and the flavor problem is a fascinating aspect of particle physics, where theoretical models seek to explain observed phenomena at the quantum level. In particular, addressing the hierarchical structure and mixing patterns in fermion masses beyond the Standard Model (SM) presents a prominent challenge in particle physics. Non-Abelian flavor symmetries play a crucial role in addressing the flavor problem, and their use in model building became common after the observation of large leptonic mixing angles by neutrino oscillation experiments \cite{R1,R2,R3}. The SSB of these discrete symmetries is the most important ingredient to achieve realistic fermion masses and mixing angles at low energies. It is achieved when gauge singlet flavon fields, which transform non-trivially under the flavor symmetry, acquire vacuum expectations values (VEVs) along specific directions in flavon space. 

On the other hand, the SSB of discrete symmetries leads to the formation of two-dimensional defects, called domain walls (DWs) with potential cosmological implications \cite{zel,kib}. These DWs form at the boundaries of distinct degenerate vacua created during the phase transition associated with SSB. In the context of non-Abelian discrete symmetries, SSB typically occurs partially, resulting in residual subgroups within different fermion sectors due to the VEV alignments of flavon fields. The number of degenerate vacua is determined by the order of the broken subgroups, while the regions between these vacua represent the DWs whose energy density may conflict with cosmological observations if it dominates the total energy density of the Universe. This happens in the so-called scaling regime in which the energy density of DWs scales in terms of the cosmological expansion factor $a(t)$ as $a^{-1}$ \cite{R4,R5,R6}, decreasing more slowly than radiation (scaling as $a^{-4}$) or matter (scaling as $a^{-3}$) \cite{zel,kib,B3}. Basically, if the flavor-breaking scale is comparable to the inflationary scale, it is reasonable to assume that the Universe has expanded enough for the walls to be inflated beyond the present horizon. However, if the flavor-breaking scale is lower than the inflationary scale, the energy density of the walls becomes the subdominant contribution to the total energy density of the Universe \cite{R7,R8}; this is known as the DW problem \cite{zel}.

In this paper, based on our recent works \cite{A4,S4}, we investigate the predictions of neutrino masses and mixing within the framework of a flavored next-to-minimal supersymmetric Standard Model (FNMSSM), where the flavor symmetry is given by $G_F = A_4 \times Z_3$. The motivation behind the NMSSM comes from the consequences associated with its singlet superfield, see Refs. \cite{R9,R10,R11} for more details. In our scenario, to achieve realistic light neutrino masses through the type I seesaw mechanism, three right-handed neutrinos and two flavon fields are introduced. Meanwhile, the NMSSM singlet $\mathcal{S}$ plays a crucial role in achieving neutrino mixing that aligns with the Trimaximal mixing matrix \cite{tm1,tm2,tm3,tm4,tm5,tm6,tm7}. We have studied numerically the phenomenology associated with neutrino sector in the normal mass hierarchy (NH) case. We found that upcoming experiments can test our predictions for the effective Majorana mass $m_{\beta\beta}$ relevant for neutrinoless double beta decay experiments ($0\nu\beta\beta$), the effective mass of electron antineutrinos $m_\beta$ measured in beta decay experiments, and the total mass of the three active neutrinos constrained by cosmological observations. 

Furthermore, we investigate the partial breaking of $A_4$, and we focus on patterns with two different residual symmetries\footnote{The extra $Z_3$ discrete group differ from the $Z_3$ subgroups of the $A_4$ symmetry, see Ref. \cite{A4} for more details.}: $A_4 \rightarrow Z_3$ in the charged lepton sector and $A_4 \rightarrow Z_2$ in the neutrino sector. For the first breaking pattern, the broken part of $A_4$ is given by the Klein four group $Z_2 \times Z_2$ of order four. In this case, DWs manifest as boundaries that separate four degenerate vacua. For the second breaking pattern, the broken part of $A_4$ is given by the non-Abelian group $Z_2 \times Z_3$ of order six. In this case, DWs manifest as boundaries that separate six degenerate vacua. By representing the degenerate vacua as vectors in flavon space, we observe that DW networks in both sectors can be represented by Platonic solids: a Tetrahedron for $A_4\rightarrow Z_3$ and an Octahedron for $A_4 \rightarrow Z_2$. In the charged lepton sector, DWs emerge around the inflationary scale without conflicting with cosmological observations. However, in the neutrino sector, the breaking occurs below the inflationary scale, requiring the annihilation of corresponding DWs before the epoch of big bang nucleosynthesis according to standard cosmology. Following Ref. \cite{zel}, we show that explicitly breaking the $Z_2 \rtimes Z_3$ subgroup of $A_4$ at high energy using Planck-suppressed higher-dimensional operators can eliminate the degeneracy among the six degenerate vacua. Notably, the low-energy behavior of the theory remains mostly unchanged as long as these operators are suppressed by powers of the Planck scale.
\section{Implementing $A_4$ in the lepton sector of the FNMSSM}
In this section, we examine the neutrino masses and mixing within the framework of the {\it FNMSSM} where the flavor symmetry is given by $G_f = A_4 \times \mathbf{Z_3}$. In order to account for neutrino oscillation data, the $A_4$ group must undergo spontaneous breaking, resulting in the emergence of cosmic domain walls. The $\mathbf{Z_3}$ symmetry, on the other hand, is primarily introduced to distinguish between the flavon superfields used in the charged lepton and neutrino sectors. $A_4$ has four irreducible representations: one triplet and three singlets which we denote here by their basis characters as $\mathbf{3}_{(-1,0)}$,  $\mathbf{1}_{(1,1)}$, $\mathbf{1}_{(1,\omega)}$ and $\mathbf{1}_{(1,\omega^2)}$. This notation provides a nuanced approach for differentiating the three one-dimensional representations by the characters of the two generators of $A_4$ (see Ref. \cite{A42} and appendix for more details). In order to prevent any confusion between the characters associated with the additional $\mathbf{Z_3}$ and the $Z_3$ subgroup of $A_4$, we represent the three one-dimensional representations of the former as $\mathbf{1_1}$, $\mathbf{1_Q}$ and $\mathbf{1_{Q^2}}$, where $\mathbf{Q} = e^{2 \pi i / 3}$.

By focusing on the lepton sector, our {\it FNMSSM} prototype incorporates, in addition to the conventional superfields of the MSSM and the NMSSM singlet $\mathcal{S}$ the following additional superfields: \textit(1) three right handed neutrinos $N_i^c = (N_1^c,N_2^c,N_3^c)$ essential for generating neutrino masses through the type-I seesaw mechanism, and \textit(2) three flavon superfields denoted as $\Phi, \Omega$, and $\chi$. These flavon are necessary for generating appropriate neutrino masses and mixing, as well as contributing to the resolution of the DW problem. The quantum numbers for both the additional fields and the MSSM fields under the $SU(2)_L \times U(1)_Y$ and the $A_4 \times \mathbf{Z_3}$ symmetries are detailed in Table \ref{t1}.
\begin{table}[h!]
\tbl{Superfields and transformation properties under $SU(2)_L \times U(1)_Y$ group and the $A_4 \times \mathbf{Z_3}$ flavor symmetry.
\label{t1}}
{\begin{tabular}{crrrrrrrrrrr}
\toprule
 Superfields & \multicolumn{1}{c}{$L_i$} & \multicolumn{1}{c}{$e^c$} & \multicolumn{1}{c}{$\mu^c$} & \multicolumn{1}{c}{$\tau^c$} & \multicolumn{1}{c}{$N_i^c$} & \multicolumn{1}{c}{$H_u$} & \multicolumn{1}{c}{$H_d$} & \multicolumn{1}{c}{$\Phi$} & \multicolumn{1}{c}{$\Omega$} & \multicolumn{1}{c}{$\mathcal{S}$} & \multicolumn{1}{c}{$\chi$}\\
\colrule
 $SU(2)_L \times U(1)_Y$ & (2,-1) & (1,2) & (1,2) & (1,2) & (1,0) & (2,1) & (2,-1) & (1,0) & (1,0) & (1,0) & (1,0)\\
 $A_4$ & $\mathbf{3}_{(-1,0)}$ & $\mathbf{1}_{(1,\omega^2)}$ & $\mathbf{1}_{(1,\omega)}$ & $\mathbf{1}_{(1,1)}$ & $\mathbf{3}_{(-1,0)}$ & $\mathbf{1}_{(1,1)}$ & $\mathbf{1}_{(1,\omega)}$ & $\mathbf{3}_{(-1,0)}$ & $\mathbf{3}_{(-1,0)}$ & $\mathbf{1}_{(1,\omega^2)}$ & $\mathbf{1}_{(1,1)}$ \\
 $\mathbf{Z_3}$ & $\mathbf{1}_Q$ & $\mathbf{1}_Q$ & $\mathbf{1}_Q$ & $\mathbf{1}_Q$ & $\mathbf{1}_{Q^2}$ & $\mathbf{1}_1$ & $\mathbf{1}_Q$ & $\mathbf{1}_1$ & $\mathbf{1}_{Q^2}$ & $\mathbf{1}_{Q^2}$ & $\mathbf{1}_{Q^2}$ \\
 \botrule
\end{tabular}}
\end{table}
By using these field transformations, the most general chiral superpotential for leptons invariant under $SU(2)_L \times U(1)_Y \times G_f$ is given by
\begin{eqnarray}
    W_Y = \frac{y^{ijk}}{\Lambda} L_i E_j H_d \Phi_k + y^{ij} L_i N_j^c H_u + \lambda_{\chi}~\chi N^c N^c + \lambda_{\Omega}~\Omega N^c N^c + \lambda_{\mathcal{S}}~\mathcal{S} N^c N^c 
\label{wy}    
\end{eqnarray}
where $L_i$ denotes the three left handed lepton doublets, $E_j^c=(e^c,\mu^c,\tau^c)$ is the three right handed charged leptons, and $H_u$ and $H_d$ are the two Higgs doublets of the MSSM. The first term in Eq. \ref{wy} is responsible for the charged lepton masses, the second term gives rise to the Dirac mass matrix, while the remaining terms are responsible for RH neutrino masses. Following the breaking of gauge and flavor symmetries, the masses of charged leptons and neutrinos are obtained through their interactions with the scalar components of the scalar superfields. The specific VEV alignments that adhere to neutrino data and minimize the scalar potential of the model are given by
\begin{eqnarray}
    \left\langle \Phi \right\rangle = \upsilon_\Phi (1,0,0), \quad \left\langle \Omega \right\rangle = \upsilon_\Omega (1,1,1), \quad \left\langle \chi \right\rangle = \upsilon_\chi, \quad \left\langle H_{u,d} \right\rangle = \upsilon_{u,d}, \quad \left\langle \mathcal{S} \right\rangle = \upsilon_\mathcal{S}.
\end{eqnarray}
Then, according to the superfield transformations in Table \ref{t1} and the $A_4$ tensor product rules given in the appendix, the charged leptons, Dirac and RH neutrinos mass matrices are given respectively as follows\footnote{Further details regarding the derivation of these matrices can be found in Ref. \cite{A4}.}
\begin{equation}
    M_l = \frac{\upsilon_d \upsilon_\Phi}{\Lambda}
    \begin{pmatrix}
        y_e & 0 & 0 \\
        0  & y_\mu & 0 \\
        0  & 0  & y_\tau
    \end{pmatrix}, \quad 
    M_D = Y_0 \upsilon_u
    \begin{pmatrix}
        1 & 0 & 0 \\
        0  & 0 & 1 \\
        0  & 1  & 0
    \end{pmatrix}, \quad
    M_R = 
    \begin{pmatrix}
        a + \frac{2b}{3} & -\frac{b}{3} + \epsilon & -\frac{b}{3} \\
        -\frac{b}{3} + \epsilon  & \frac{2b}{3} & a -\frac{b}{3} \\
        -\frac{b}{3}  & a -\frac{b}{3}  & \frac{2b}{3} + \epsilon
    \end{pmatrix},
\label{ldr}    
\end{equation}
where $a = 2 \lambda_\chi \upsilon_\chi$, $b = 2 \lambda_\Omega \upsilon_\Omega$, and $\epsilon = 2 \lambda_\mathcal{S} \upsilon_\mathcal{S}$. The obtained mass matrix for charged leptons is diagonal with the mass eigenvalues given as
\begin{eqnarray}
    m_e = y_e \frac{\upsilon_d \upsilon_\Phi}{\Lambda}, \quad m_\mu = y_\mu \frac{\upsilon_d \upsilon_\Phi}{\Lambda}, \quad m_\tau = y_\tau \frac{\upsilon_d \upsilon_\Phi}{\Lambda}.
    \label{ml+}
\end{eqnarray}
For neutrinos, the Majorana mass matrix $M_R$ exhibits the magic symmetry referring to the fact that the sum of each column and row is equal \cite{ms}. This type of matrices are diagonalized by the well-known Trimaximal mixing matrix defined as \cite{tm1,tm2,tm3,tm4,tm5,tm6,tm7}
\begin{equation}
    U_{TM_2} = 
    \begin{pmatrix}
        \sqrt{\frac{2}{3}} \cos\theta & \frac{1}{\sqrt 3} & \sqrt{\frac{2}{3}} \sin\theta e^{-i \sigma} \\
        \frac{-\cos\theta}{\sqrt{6}} - \frac{\sin\theta}{\sqrt{2}} e^{i \sigma} & \frac{1}{\sqrt 3} & \frac{\cos\theta}{\sqrt{2}} - \frac{\sin\theta}{\sqrt{6}} e^{-i \sigma} \\
        \frac{-\cos\theta}{\sqrt{6}} + \frac{\sin\theta}{\sqrt{2}} e^{i \sigma} & \frac{1}{\sqrt 3} & \frac{-\cos\theta}{\sqrt{2}} - \frac{\sin\theta}{\sqrt{6}} e^{-i \sigma}
    \end{pmatrix}.
    \label{TM2}
\end{equation}
The parameters $a, b$ and $\epsilon$ in $M_R$ have a dimension of mass and at least one of them must be complex in order to account for {\it CP} violation in the lepton sector. The optimal choice can be deduced by noticing that when $\epsilon \rightarrow 0$, $M_R$ is diagonalized by the tribimaximal mixing (TBM) matrix which is known to be incompatible with the present neutrino oscillation data and it is {\it CP} conserving \cite{tbm}. Therefore, $\epsilon$ is responsible for a small deviation of the neutrino mixing angles from their TBM values and it is taken to be complex ($\epsilon = |\epsilon|e^{i \phi_\epsilon}$) to account for the {\it CP} violation in the neutrino sector. Moreover, since this parameter is proportional to the VEV of the NMSSM singlet $\mathcal{S}$ known to acquire a VEV of the order of $M_{SUSY}$, the modulus of $\epsilon$ must satisfy: $|\epsilon| < a,b$. Now, let us proceed to compute the masses, starting with the three Majorana masses  $M_{i=1,2,3}$. These masses are determined through the diagonalization of the right-handed neutrino mass matrix $M_R$ using the Trimaximal mixing matrix as $(U_{TM_2}^\ast)^T M_R  U_{TM_2}^\ast = diag(M_1,M_2,M_3) $ where
\begin{equation}
    M_1 = a + b - \frac{\left\lvert \epsilon \right\rvert}{2} + O\left(\frac{\left\lvert \epsilon \right\rvert^2}{a^2}\right), \quad M_2 = a + \left\lvert \epsilon \right\rvert, \quad M_3 = b - a + \frac{\left\lvert \epsilon \right\rvert}{2} + O\left(\frac{\left\lvert \epsilon \right\rvert^2}{a^2}\right).
    \label{maj}
\end{equation}
The diagonalization of $M_R$ is performed under the fulfillment of the following relations between the model parameters \{$a$, $b$, $\lvert \epsilon \rvert$, $\phi_\epsilon$ \} and the Trimaximal mixing parameters $\theta$ and $\sigma$
\begin{equation}
    \tan2\theta = \frac{\lvert \epsilon \rvert \sqrt{3 b^2 \cos\phi^2_\epsilon + 3 a^2 \sin\phi^2_\epsilon}}{b \lvert \epsilon \rvert \cos\phi_\epsilon - 2 b a}, \quad \tan\sigma = \frac{a}{b}\tan\phi_\epsilon .
\end{equation}
At this point, we can calculate the light neutrino masses via the type-I seesaw mechanism formula $m_\nu = m_D^T M_R^{-1} m_D$. Given that $M_R$ is diagonalized by $U_{TM_2}$ given in Eq. \ref{TM2}, the inverse Majorana neutrino mass is expressed as 
$M_R^{-1} = U_{TM_2}^\ast [diag(M_1, M_2, M_3)]^{-1} (U_{TM_2}^\ast)^T$. Moreover, due to the form of the Dirac mass matrix $M_D$, see Eq. \ref{ldr}, the diagonalization of $m_\nu$ remains of Trimaximal form, but with the interchange of the second and the third row of $U_{TM_2}$. We shall denote the new form of the Trimaximal mixing matrix as $\Tilde{U}_{TM_2}$ and thus, the light neutrino masses are given by $(\Tilde{U}_{TM_2})^T m_\nu \Tilde{U}_{TM_2} = \textbf{diag}(m_1,m_2,m_3)$ where $m_i = Y_0^2 \upsilon_u^2 / M_i$. To make our upcoming numerical study more straightforward, we redefine these masses as a function of $O(1)$ parameters as follows
\begin{equation}
    m_1 = \frac{2 m_0}{2 + 2 k_{ab} - k_{\epsilon b}}, \quad m_2 = \frac{m_0}{k_{a b} + k_{\epsilon b}}, \quad m_3 = \frac{2 m_0}{2 - 2 k_{a b} + k_{\epsilon b} },
    \label{ev}
\end{equation}
where $m_0 = Y_0^2 \upsilon_u^2 / b$, $k_{ab} = a/b$ and $k_{\epsilon b} = \left\lvert \epsilon \right\rvert / b$. Regarding the neutrino mixing angles, we can easily derive their expressions from their respective definitions: $\sin^2\theta_{13} = \lvert \Tilde{U}_{e3} \rvert^2$, $\sin^2\theta_{12} = \lvert \Tilde{U}_{e2} \rvert^2 / (1 - \lvert \Tilde{U}_{e3} \rvert^2)$, and $\sin^2\theta_{23} = \lvert \Tilde{U}_{\mu 3} \rvert^2 / (1 - \lvert \Tilde{U}_{e3} \rvert^2)$, where $\Tilde{U}_{e2}$, $\Tilde{U}_{e3}$ and $\Tilde{U}_{\mu 3}$ represent the (12), (13) and (23) entries of the Trimaximal mixing matrix $\Tilde{U}_{TM_2}$, respectively. Thus, we conclude that the expression for these mixing angles are given by
\begin{equation}
    \sin^2\theta_{13} = \frac{2}{3} \sin^2\theta, \quad \sin^2\theta_{12} = \frac{1}{3 - 2 \sin^2\theta}, \quad \sin^2\theta_{23} = \frac{1}{2} + \frac{\sqrt{3} \sin2\theta}{2(3 - 2 \sin^2\theta)} \cos\sigma.
    \label{mix}
\end{equation}
\section{Domain walls from spontaneous breaking of $A_4$}
In this section, we examine the formation of DWs and depict them geometrically in the context of two $A_4$ breaking patterns, i.e. $A_4 \rightarrow Z_3$, $A_4 \rightarrow Z_2$. To simplify things, we use the fact that $A_4$ is isomorphic to $(Z_{2}\times Z_{2})\rtimes Z_{3} $, where $(Z_{2}\times Z_{2})$ is the Klein four-group which will be referred to as $K_{4}$ in subsequent discussions. Let us start by delving into the different breaking patterns within the basis provided in equation \ref{basis} of the appendix:
\begin{itemize}
    \item $\mathbf{A_4 \rightarrow Z_3}$: Let us denote the isomorphic group of $A_4$ as $A_4 \cong Z_2^S \times Z_2^{TST^2} \rtimes Z_3^T$ where the exponents represent the generator for each abelian subgroup of $A_4$. Accordingly, the breaking pattern $A_4 \rightarrow Z_3^T$ is realized in the charged leptons sector when the flavon triplet $\Phi$ acquires its VEV along the direction $\left\langle \Phi \right\rangle = (1,0,0)^T$. This can be easily checked using the matrix representations for the generators $S$ and $T$ provided in the appendix where we find
    \begin{equation}
        S \left\langle \Phi \right\rangle \neq \left\langle \Phi \right\rangle, \quad TST^2 \left\langle \Phi \right\rangle \neq \left\langle \Phi \right\rangle, \quad T \left\langle \Phi \right\rangle = \left\langle \Phi \right\rangle. 
    \end{equation}
    This equation makes it clear that among the symmetries, only $Z_3^T$ remains unbroken. On the other hand, the part that undergoes breaking is given by the Klein four group $K_4=Z_2^S \times Z_2^{TST^2}$. It is crucial to emphasize that choosing the third $Z_2^{T^2ST}$ subgroup within $A_4$ instead of either $Z_2^S$ or $Z_2^{TST^2}$ yields to the same outcomes. This arise because the generator $T^2ST$ of this group does not preserve the VEV of $\Phi$, i.e. $T^2ST \left\langle \Phi \right\rangle \neq \left\langle \Phi \right\rangle$. On the other hand, we cannot choose the other three $Z_3$ subgroups within $A_4$--namely $Z_3^{ST}, Z_3^{TS} \text{and } Z_3^{STS}$--instead of $Z_3^T$. This is because they all modify the VEV structure of $\Phi$ leading to the breaking of the $Z_3$ symmetry, i.e. $ST \left\langle \Phi \right\rangle \neq \left\langle \Phi \right\rangle$, $TS \left\langle \Phi \right\rangle \neq \left\langle \Phi \right\rangle$ and $STS \left\langle \Phi \right\rangle \neq \left\langle \Phi \right\rangle$.
    \item $\mathbf{A_4 \rightarrow Z_2}$: We employ the same generators for the isomorphic group as in the previous breaking pattern. Let us assume that the residual symmetry is the $Z_2^S$ group. Thus, the breaking pattern $A_4 \rightarrow Z_2^S$ in the neutrino sector is realized when the flavon triplet $\Omega$ acquires its VEV along the direction $\left\langle \Omega \right\rangle = (1,1,1)^T$. As in the previous case, we can verify this breaking pattern using the matrix representations for the generators $S$ and $T$, and show that the only conserved symmetry is the that of the generator $S$. We have
        \begin{equation}
        S \left\langle \Omega \right\rangle = \left\langle \Omega \right\rangle, \quad TST^2 \left\langle \Omega \right\rangle \neq \left\langle \Omega \right\rangle, \quad T \left\langle \Omega \right\rangle \neq \left\langle \Omega \right\rangle. 
    \end{equation}
    This breaking pattern does not occur when considering the residual group as either of the remaining two $Z_2$ subgroups within $A_4$, namely $Z_2^{TST^2}$ and $Z_2^{T^2ST}$. This restriction arises from the fact that both of these subgroups undergo breaking due to the VEV direction of $\Omega$. Conversely, choosing any of the alternate $Z_3$ subgroups within $A_4$ instead of $Z_3^T$ poses no issue, as they are also broken by the VEV of $\Omega$. Specifically, conditions such as $ST \left\langle \Omega \right\rangle \neq \left\langle \Omega \right\rangle$, $TS \left\langle \Omega \right\rangle \neq \left\langle \Omega \right\rangle$, and $STS \left\langle \Omega \right\rangle \neq \left\langle \Omega \right\rangle$ are satisfied.
    \item $\mathbf{A_4 \rightarrow K_4}$: The $A_4$ group has only one $K_4$ subgroup generated by $\{ S, TST^2\}$. As we showed in the previous breaking patterns, the action of $TST^2$ on $\left\langle \Phi \right\rangle$ and $\left\langle \Omega\right\rangle$ modifies the VEV directions of the flavon triplets. Consequently, these flavon triplets cannot be employed for the breaking pattern $A_4 \rightarrow K_4$. On the other hand, this breaking pattern can be realized by any scalar field transforming as a nontrivial singlet. Indeed, the VEV of the NMSSM singlet $\mathcal{S}$ transforming as $\mathbf{1}_{(1,\omega^2)}$ remains unchanged under the operations of the generators of $K_4$ ($S$ and $TST^2$). As a result, $\left\langle \mathcal{S} \right\rangle$ breaks $A_4$ down to its Klein four group $K_4$. In this case, the broken part corresponds to the abelian group $Z_3$, and it can be chosen from any of the four $Z_3$ subgroups within $A_4$, given that their generators all modify the VEV of the NMSSM singlet, i.e. $T \left\langle \mathcal{S} \right\rangle \neq \left\langle \mathcal{S} \right\rangle$, $ST \left\langle \mathcal{S} \right\rangle \neq \left\langle \mathcal{S} \right\rangle$, $TS \left\langle \mathcal{S} \right\rangle \neq \left\langle \mathcal{S} \right\rangle$, and $STS \left\langle \mathcal{S} \right\rangle \neq \left\langle \mathcal{S} \right\rangle$.
     It is important to note that this discussion is purely based on mathematical considerations and that these four $Z_3$ subgroups of $A_4$ are conjugate to each other. However, once a generator is chosen for a subgroup, it must be consistently maintained throughout the analysis.
\end{itemize}
Now, we examine the induced DWs within both the charged lepton and neutrino sectors, illustrating a nuanced approach to geometrically represent them for each breaking pattern. Given that these breaking patterns emerge when flavon fields attain VEVs along particular directions within the flavon space, it is essential to accurately define this space in which we will depict the DWs. Since there are two possible dimensions for the irreducible representations of $A_4$, our approach revolves around visualizing the flavon space as a (3+1)-dimensional vector space denoted by $V$ in what follows. Here, $3$ and $1$ represent the dimensions of the triplet and the singlets of $A_4$, respectively. Consider the components of this vector space labeled as $V = (X_1,X_2,X_3,X_4) \in \mathbb{C}^4$ where the $X_i$’s stand for a system of vector basis with the first three components are reserved for $A_4$ triplets, while the forth one is designated for $A_4$ singlets. Thus, the nontrivial scalar superfields $\Phi$ and $\Omega$ along with the NMSSM singlet $\mathcal{S}$ are expanded within the flavon space in terms of the components of the complex {\it 4D} vector space $V$ as\footnote{Generally, we can use a collective flavon superfield $\phi \sim \phi_k X_k$ where $k=1,2,3,4$. Hence, we may think of the linear combinations of these $X's$ components as forming the basis of the the flavon space.} $\Phi \sim \Phi_i X_i$, $\Omega \sim \Omega_i X_i$, and $\mathcal{S} = \mathcal{S} X_4$ where $i=1,2,3$. Representing DWs graphically becomes a challenging task due to the complex nature of these superfields and their scalar components that acquire VEVs. Hence, to represent the DWs by real geometrical objects such as real quivers, we adopt the methodology outlined in Refs. \cite{A4,S4}. This relies on splitting the components of the scalar superfields and the vector space $V$ into real and imaginary parts as $\phi_k = \mathbf{\Re}(\phi_{k}) + i\mathbf{\Im} (\phi_{k})$ and $V_{k} = \mathbf{\Re} (U_{k}) + i\mathbf{\Im} (R_{k})$ where $k=1,2,3,4$ while $U_k$ and $V_k$ are real {\it 4D} vectors forming the constituents of $X_k$ expressed as $X_k=(U_k,R_k)^T$. Accordingly, the complex {\it 4D} expansion $\phi \sim \phi_k X_k$ is expressed with this splitting as a real {\it 8D} vector given by
\begin{equation}
    \phi \sim \sum_k [ \mathbf{\Re}(\phi_k)U_k + \mathbf{\Im}(\phi_k)R_k ] \in \mathbb{R}^8,\quad \text{with} \quad k=1,2,3,4.
    \label{crb}
\end{equation}
Let us elaborate on this by considering a specific example of a VEV direction with complex entries. Assuming a breaking pattern of $A_4 \rightarrow Z_3^{TS}$. Given the generator $TS$ of $Z_3$, this breaking occurs when a flavon triplet $\varphi$, attains a VEV along the direction $\left\langle \varphi \right\rangle = \upsilon_\varphi (1, -2\omega, -2\omega^2)^T$ where $\omega = -1/2 + i \sqrt{3}/2$. In this scenario, the representation of $\left\langle \varphi \right\rangle$ in the real basis is expressed as {\it 6D} vector in $\mathbb{R}^6$
\begin{equation}
    \left\langle \varphi \right\rangle = \upsilon_\varphi (1,0,1,-\sqrt{3},1,\sqrt{3}).
    \label{rb}
\end{equation} 
Geometrically, the flavon VEV responsible for the breaking of $A_4$ can be represented as a vector aligned along one of the real directions within the vector space ($U_k$, $R_k$) where the components of the vector correspond to its coordinates within this space. Then, the transformations imposed on the flavon VEV by the elements of the $A_4$ group result in the establishment of coordinates for the remaining vertices, eventually shaping a polygon within the vector space ($U_k$, $R_k$). The number of vertices equals to the number of degenerate vacua, i.e. the order of the broken group. Let us explore this in our model for the breaking patterns realized in the charged lepton and neutrino sectors. 
\begin{figure}[thb]
\centering
\includegraphics[width=0.27\linewidth]{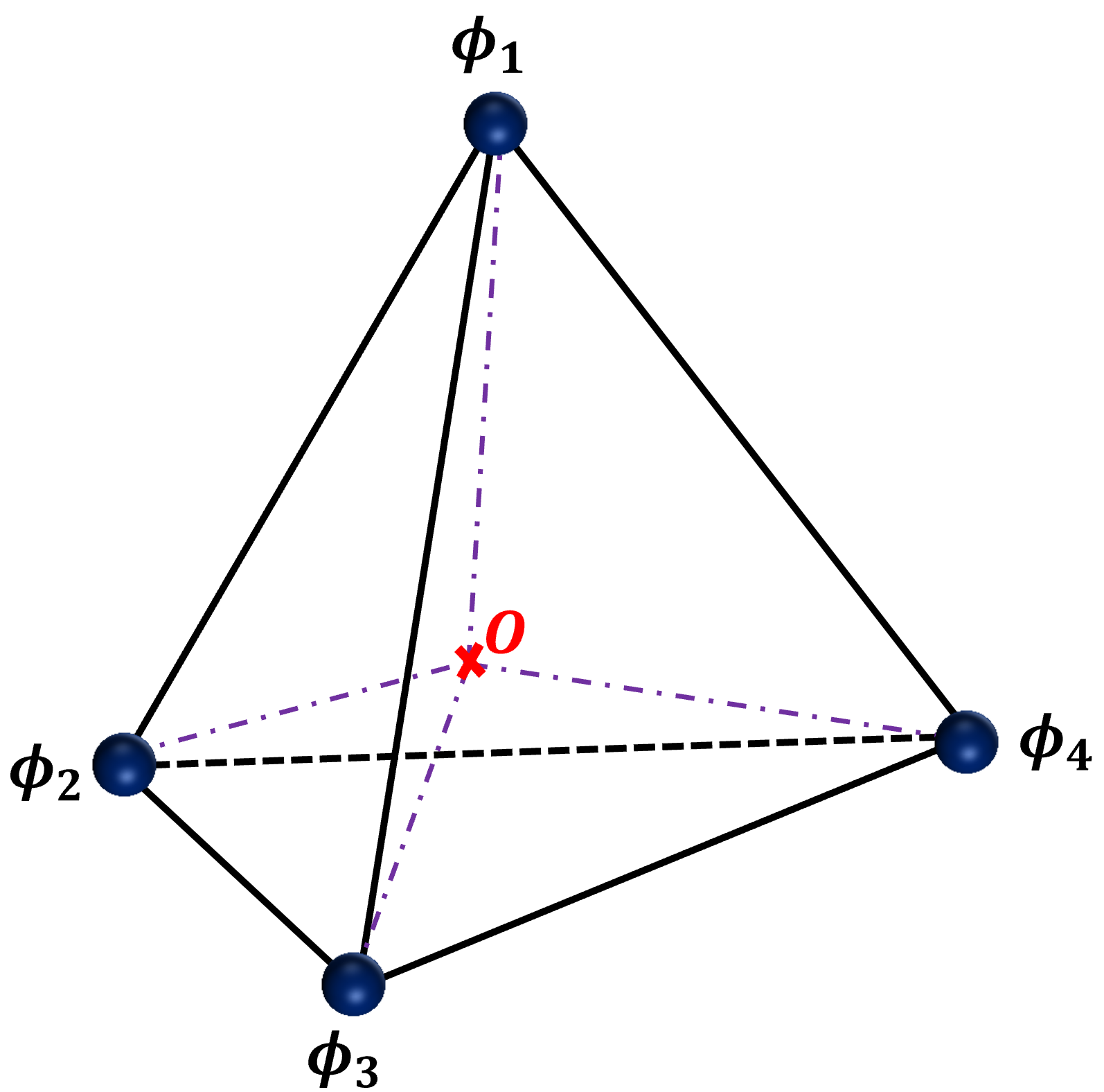}
 \hspace{2cm}
\includegraphics[width=0.27\linewidth]{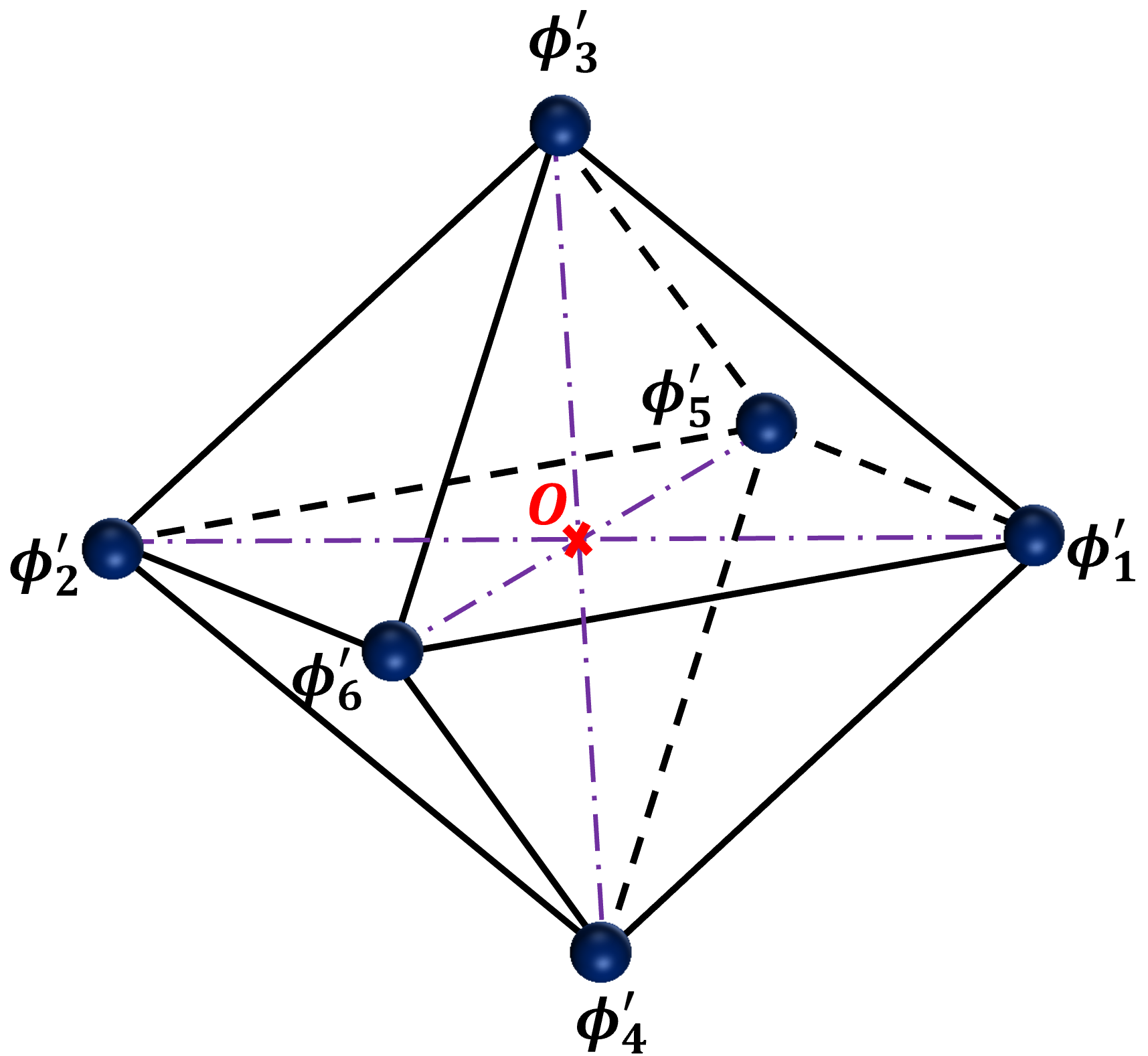}
\caption{A regular Tetrahedron (left) and a regular Octahedron (right) viewed from the flavon space. The vertices represent the degenerate vacua, while the edges that connect them stand for DWs. In these Platonic solids, the central point $O$ reflects the complete $A_4$ symmetry.}
\label{dws}
\end{figure}
As explained previously, the breaking pattern $A_4 \rightarrow Z_3^T$ manifests in the charged leptons sector when the flavon triplet $\Phi$ obtains a VEV along the direction $\left\langle \Phi \right\rangle = \upsilon_\Phi(1,0,0)^T$. This breaking scenario can be represented as $Z_2^S \times Z_2^{TST^2} \rtimes Z_3^T \rightarrow Z_3^T$, where the broken subgroup $Z_2^S \times Z_2^{TST^2}$ is isomorphic to the Klein four group $K_4$ with an order of four. Consequently, there exist four degenerate vacua $\{\phi_1,\phi_2,\phi_3,\phi_4 \}$ associated with this breaking pattern, situated within the flavon space. These vacua collectively define the vertices of a Platonic solid, specifically a Tetrahedron, as depicted in the left panel of Fig. \ref{dws}. One of the vertices of the Tetrahedron corresponds to the $K_4$-invariant vacuum employed in the $A_4 \rightarrow Z_3^T$ breaking. To derive the remaining three vacua, we apply the generators of $K_4$ to $\left\langle \Phi \right\rangle$ as follows
\begin{equation}
    \phi_1 = \left\langle \Phi \right\rangle, \quad \phi_2 = S\left\langle \Phi \right\rangle, \quad \phi_3 = TST^2\left\langle \Phi \right\rangle, \quad \phi_4 = T^2ST\left\langle \Phi \right\rangle,
\end{equation}
Using the matrix representations of the generators $S$ and $T$ in Eq. \ref{basis} of the appendix, it is easy to deduce the expressions of these vacua in the complex three-dimensional space $\mathbb{C}^3$
\begin{equation}
    \phi_1 = \upsilon_\Phi
    \begin{pmatrix}
        1 \\ 0 \\ 0
    \end{pmatrix}, \quad
    \phi_2 = \frac{\upsilon_\Phi}{3}
    \begin{pmatrix}
        -1 \\ 2 \\ 2
    \end{pmatrix}, \quad
    \phi_3= \frac{\upsilon_\Phi}{3}
    \begin{pmatrix}
        -1 \\ 2 \omega^2 \\ 2 \omega
    \end{pmatrix}, \quad
    \phi_4 = \frac{\upsilon_\Phi}{3}
    \begin{pmatrix}
        -1 \\ 2 \omega \\ 2 \omega^2
    \end{pmatrix}.
    \label{phi}
\end{equation}
In the real vector basis \ref{crb}, these vacua are expressed following the same procedure used to obtain Eq. \ref{rb}. Therefore, for the triplets $\phi_i$ we have
\begin{equation}
    \phi_1 = \upsilon_\Phi(1,0,0,0,0,0)^T, \quad \phi_2 = \frac{\upsilon_\Phi}{3}(-1,0,2,0,2,0)^T, \quad \phi_3 = \frac{\upsilon_\Phi}{3}(-1,0,-1,-\sqrt{3},-1,\sqrt{3})^T, \quad \phi_4 = \frac{\upsilon_\Phi}{3}(-1,0,-1,\sqrt{3},-1,-\sqrt{3})^T.
    \label{tetra}
\end{equation}
These four vectors satisfy the constraint $\sum_{i-1}^4\phi_i=0$, a property that characterizes a Tetrahedron with 4 vertices, 4 faces and 6 edges. These edges represent the DWs separating the vacua as illustrated in the left panel of Fig. \ref{dws}. Moreover, by examining these vectors, it is clear that the structure resembles a regular Tetrahedron where all sides form equilateral triangles, and we have confirmed that all edges share the same length, specifically $\upsilon_\Phi \sqrt{8/3}$.

In the neutrino sector, the breaking pattern $A_4 \rightarrow Z_2^S$ is realized when the flavon triplet $\Omega$ acquires a VEV along the direction $\left\langle \Omega \right\rangle = \upsilon_\Phi(1,1,1)^T$. This breaking can be expressed as $Z_2^S \times Z_2^{TST^2} \rtimes Z_3^T \rightarrow Z_2^S$, where the broken part of $A_4$ is given by the non-Abelian group $Z_2^{TST^2} \rtimes Z_3^T$ with an order of six. Consequently, six degenerate vacua $\{ {\phi^{\prime}_1,\phi^{\prime}_2,\phi^{\prime}_3,\phi^{\prime}_4,\phi^{\prime}_5,\phi^{\prime}_6 } \}$ are associated with this breaking pattern and are fixed in the flavon space. These vacua define the vertices of a Platonic solid, specifically a regular Octahedron, as depicted in the right panel of Fig. \ref{dws}. One of the vertices of this Octahedron corresponds to the $Z_2^{TST^2} \rtimes Z_3^T$-invariant vacuum employed in the $A_4 \rightarrow Z_2^S$ breaking, $\phi^{\prime}_1 = \left\langle \Omega \right\rangle$. To derive the remaining five vacua, we apply the generators of $Z_2^{TST^2} \rtimes Z_3^T$ to $\left\langle \Omega \right\rangle$ as follows
\begin{equation}
    \phi^{\prime}_1 = \left\langle \Omega \right\rangle, \quad \phi^{\prime}_2 = TST^2\left\langle \Omega \right\rangle, \quad \phi^{\prime}_3 = T\left\langle \Omega \right\rangle, \quad \phi^{\prime}_4 = TS\left\langle \Omega \right\rangle, \quad \phi^{\prime}_5 = T^2\left\langle \Omega \right\rangle, \quad
    \phi^{\prime}_6 = TST\left\langle \Omega \right\rangle.
\end{equation}
Using the matrix representations of the generators $S$ and $T$ in Eq. \ref{basis}, the expressions of these vacua in the complex three-dimensional space $\mathbb{C}^3$ are given by
\begin{equation}
    \phi^{\prime}_1 = \upsilon_\Omega
    \begin{pmatrix}
        1 \\ 1 \\ 1
    \end{pmatrix}, \quad
    \phi^{\prime}_2 = -\upsilon_\Omega
    \begin{pmatrix}
        1 \\ 1 \\ 1
    \end{pmatrix}, \quad
    \phi^{\prime}_3= \upsilon_\Omega
    \begin{pmatrix}
        1 \\ \omega^2 \\ \omega
    \end{pmatrix}, \quad
    \phi^{\prime}_4 = -\upsilon_\Omega
    \begin{pmatrix}
        1 \\ \omega^2 \\ \omega
    \end{pmatrix}, \quad
    \phi^{\prime}_5 = \upsilon_\Omega
    \begin{pmatrix}
        1 \\ \omega \\ \omega^2
    \end{pmatrix}, \quad
    \phi^{\prime}_6 = -\upsilon_\Omega
    \begin{pmatrix}
        1 \\ \omega \\ \omega^2
    \end{pmatrix}
    \label{phipr}
\end{equation}
In the real vector basis \ref{crb}, these vacua are expressed following the same procedure used to obtain Eq. \ref{tetra}. We have
\begin{equation}
    \phi^{\prime}_{1,2} = \pm \upsilon_\Omega(1,0,1,0,1,0)^T, \quad \phi^{\prime}_{3,4} = \pm \upsilon_\Omega(1,0,-1/2,-\sqrt{3}/2,-1/2,\sqrt{3}/2)^T, \quad \phi^{\prime}_{5,6} = \pm \upsilon_\Omega(1,0,-1/2,\sqrt{3}/2,-1/2,-\sqrt{3}/2)^T.
    \label{octa}
\end{equation}
These six vectors are subject to the constraint $\sum_{i-1}^6\phi^{\prime}_i=0$, a property that characterizes an Octahedron with 6 vertices, 8 faces and 12 edges. These edges represent the DWs separating the vacua as illustrated in the right panel of Fig. \ref{dws}. Let us verify some of the properties of this regular Octahedron using the six vacua in Eq. (\ref{octa}). Consider the equal edge lengths of the regular Octahedron (right panel of Fig. \ref{dws}); for instance, in the right triangular face defined by vertices $\{ \phi^{\prime}_{1}$, $\phi^{\prime}_{3},\phi^{\prime}_{5} \}$, we find $d_{13}=d_{15}=d_{35}=\sqrt{6}\upsilon_{\Omega}$. Here, $d_{13}$, $d_{15}$, and $d_{35}$ represent the lengths of edges between $\phi^{\prime}_{1}$ and $\phi^{\prime}_{3}$, $\phi^{\prime}_{1}$ and $\phi^{\prime}_{5}$, and $\phi^{\prime}_{3}$ and $\phi^{\prime}_{5}$, respectively. Additionally, the three diagonals within the regular Octahedron are equal in the length. From the notations of Fig. \ref{dws}, we find $d_{12}=d_{34}=d_{56}=2\sqrt{3}\upsilon_{\Omega}$, where $d_{12}$, $d_{34}$, and $d_{56}$ represent the lengths of diagonals between $\phi^{\prime}_{2}$ and $\phi^{\prime}_{1}$, $\phi^{\prime}_{4}$ and $\phi^{\prime}_{3}$, and $\phi^{\prime}_{6}$ and $\phi^{\prime}_{5}$, respectively.
\section{Numerical results and solution to the domain wall problem}
This section comprises two parts. The first one is devoted to a numerical analysis of the neutrino sector, with a focus on neutrino oscillation parameters and observables associated with neutrino masses, namely $m_{\beta\beta}$, $m_\beta$, and $\sum m_i$. The second part is dedicated to presenting a solution to the DW problem.
\subsection{Neutrino phenomenology}
\begin{figure}[thb]
\centering
\includegraphics[width=0.32\linewidth]{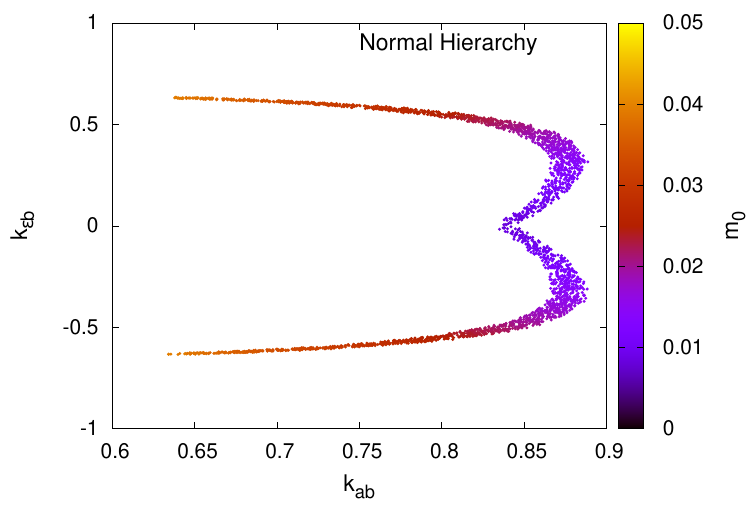}
\includegraphics[width=0.32\linewidth]{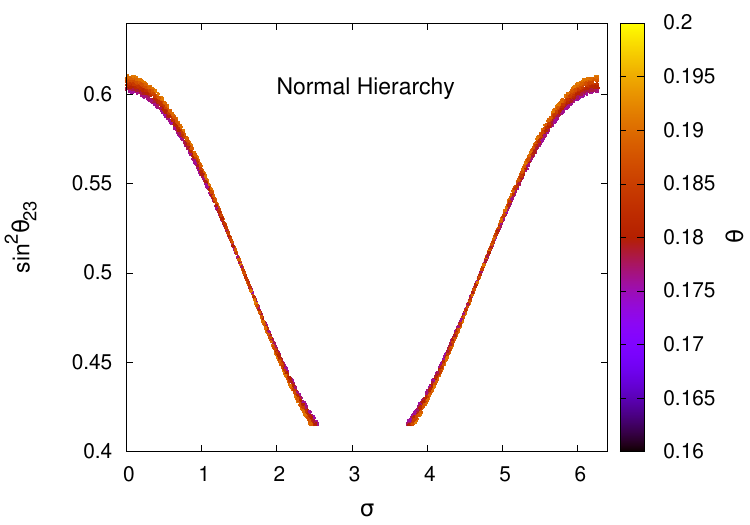}
\includegraphics[width=0.32\linewidth]{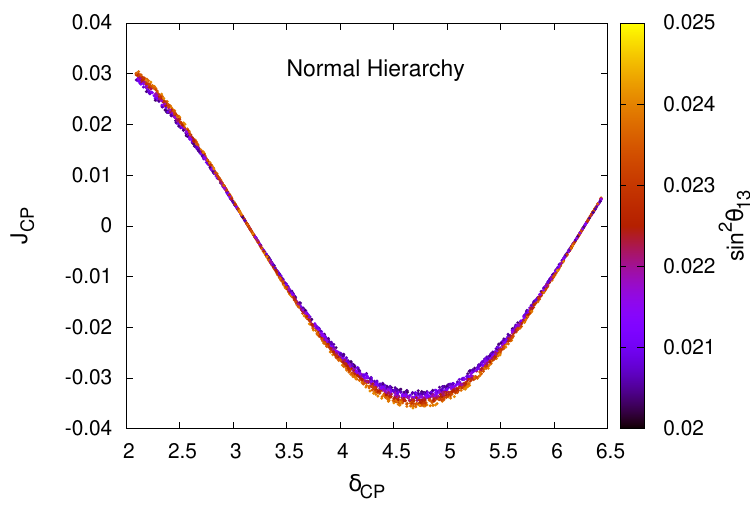}
\caption{Left: Scatter plot of $k_{ab}$ versus $k_{\epsilon b}$ with the palette showing the allowed values of $m_0$. Middle: Scatter plot of $\sin^2 \theta_{23}$ versus $\sigma$ with the palette showing the allowed values of $\theta$. Right: Scatter plot of $J_{CP}$ versus $\delta_{CP}$ with the palette showing the allowed values of $\sin^2 \theta_{13}$.}
\label{para}
\end{figure}

The current neutrino oscillation data exhibit a slight inclination towards the normal neutrino mass hierarchy with $m_1$ being the lightest neutrino mass. Consequently, our numerical investigation will be conducted within the NH framework. Let us start by fixing our model parameters given by $m_0$, $k_{ab}$, $k_{\epsilon b}$, $\theta$, $\sigma$ and $\phi_\epsilon$. The parameters $m_0$, $k_{ab}$, $k_{\epsilon b}$ are allowed to varie in the ranges $[0,1]$, $[0,1]$ and $[-1,1]$, respectively. The parameter $\theta$ is varied in the range $[0,\pi/2]$ while the phases $\sigma$ and $\phi_\epsilon$ are randomly varied in the range $[0,2\pi]$. Using the $3\sigma$ allowed range of the mixing angles from NuFit 5.0 global analysis \cite{A1} as input, and the their expressions in Eq. \ref{mix}, we can easily fix the Trimaximal parameters $\theta$ and $\sigma$. In the middle panel of Fig. \ref{para}, we depict the correlation between $\sin^2\theta_{23}$ and $\sigma$ with the color bar showing the values of $\theta$. The obtained ranges for these parameters are: $\sin^2\theta_{23} \in [0.415,0.610]$, $\theta \in [0.175,0,191]$ and $\sigma \in [0.001,6,270]$, while the entire range of $\phi_\epsilon$ is permissible. From the range of $\sin^2\theta_{23}$, we deduce that both octants of the atmospheric angle are allowed in this model.
In the left panel of Fig. \ref{para}, we show the correlation among the parameters $k_{ab}$, $k_{\epsilon b}$ and $m_0$ implicated in the expressions of neutrino masses in Eq. \ref{ev}. These parameters are constrained using the current $3\sigma$ allowed range of the mass squared differences $\Delta m_{21}^2$ and $\Delta m_{31}^2$ where we find the following ranges: $k_{ab} \in [0.634,0.888]$, $k_{\epsilon b} \in [-0.634,0.634]$ and $m_0[GeV] \in [0.007,0.039]$. Regarding the size of {\it CP} violation in the lepton sector, it can be measured using the Jarlskog invariant parameter expressed in the Particle Data Group (PDG) standard parametrization as $J_{CP} = \frac{1}{8} \sin2\theta_{12} \sin2\theta_{13} \sin2\theta_{23} \cos\theta_{13} \sin\delta_{CP}$ where $\delta_{CP}$ is the {\it CP} violating phase \cite{PDG}. In the right panel of Fig. \ref{para}, we show a scatter plot of $J_{CP}$ versus $\delta_{CP}$ while the color bar displays the $3\sigma$ allowed values of $\sin^2 \theta_{13}$. The obtained range for the Jarlskog invariant parameter is $J_{CP} \in [-0.035,0,030]$. In the case of the Trimaximal mixing matrix $\Tilde{U}_{TM_2}$, the Jarlskog parameter is expressed as $J_{CP}^{TM_2} = - (1/6\sqrt{3}) \sin{2\theta_{23}} \sin\sigma$. By matching this expression with the one provided by the PDG, a relationship emerges among the atmospheric angle, the Dirac {\it CP} phase, and $\sigma$ 
\begin{equation}
  \sin\sigma = -\sin{2\theta_{23}} \sin\delta_{CP}.
  \label{tmc}
\end{equation}
By considering this equation and the $3\sigma$ allowed ranges for the atmospheric angle and the Dirac {\it CP} phase, we find that the exact value of $n\pi$ for $\sigma$ and $\delta_{CP}$ is excluded, with $n$ can be any integer. Hence, it is evident from Eq. \ref{tmc} that the {\it CP} conserving values of $\delta_{CP}$ are not allowed in the current model, which implies also that the Jarlskog invariant parameter remains nonzero, leading to the inevitable presence of {\it CP} violation in the current model.
\begin{figure}[thb]
\centering
\includegraphics[width=0.325\linewidth]{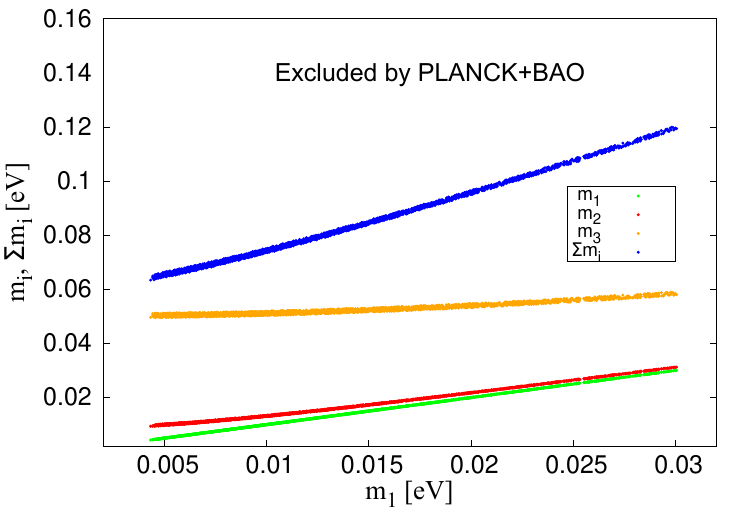}
\includegraphics[width=0.325\linewidth]{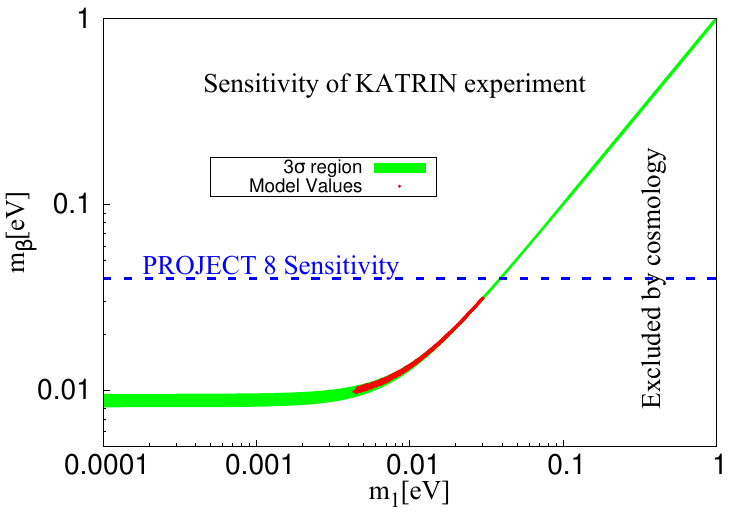}
\includegraphics[width=0.325\linewidth]{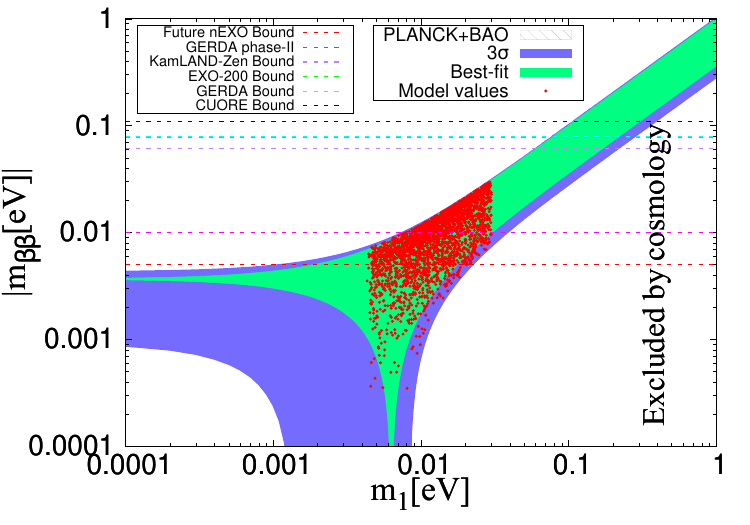}
\caption{Left: $m_i$ and $\Sigma m_i$ as a function of $m_1$. Middle: The effective neutrino mass $m_\beta$ versus $m_1$. Right: $|m_{\beta \beta|}$ versus $m_1$ where the horizontal dashed lines represent the limits on $|m_{\beta \beta}|$ from current and future $0 \nu \beta \beta$ decay experiments. The vertical and horizontal dashed regions are disfavored by experimental data.}
\label{pheno}
\end{figure}

Now, we proceed to investigate the absolute neutrino mass spectrum through scatter plots. To determine all three neutrino masses in the NH case, we represent $m_{2}$ and $m_{3}$ in terms of the lightest neutrino mass $m_{1}$, namely $m_2 = (m_{1}^{2}+\Delta m_{21}^{2})^{1/2}$ and $m_3 = (m_{1}^{2}+\Delta m_{31}^{2})^{1/2}$. Accordingly, for our numerical study, we use as input the upper bound on $\sum m_i$ from the latest Planck data given by $\sum m_i < 0.12~eV$ at 95\% Confidence Level (C.L) \cite{plck} as well as the $3\sigma$ allowed intervals for the mass squared differences $\Delta m_{21}^{2}$ and $\Delta m_{31}^{2}$ given by the NuFit 5.0 global analysis \cite{A1}. In the left panel of Fig. \ref{pheno}, we present a scatter plot depicting the three neutrino masses and their sum as a function of $m_1$. The predicted ranges of these parameters are given by
\begin{equation}
    0.004 \lesssim m_1[eV] \lesssim 0.030, \quad 0.009 \lesssim m_2[eV] \lesssim 0.031, \quad 0.049 \lesssim m_3[eV] \lesssim 0.058, \quad 0.063 \lesssim \sum m_i[eV] \lesssim 0.119.
\end{equation}
The upper value for $\sum m_i$ aligns with the upper limit reported by the Planck collaboration \cite{plck}. However, the lower value $\sim 0.063~eV$ requires additional investigations, and may be tested in future experiments, such as CORE+BAO which aim to achieve a sensitivity of $0.062~eV$ \cite{core}. There are two additional avenues for exploring the absolute mass scale of neutrinos. The first one involves a direct determination through the measurement of the electron energy spectrum near its endpoint region. This method currently stands as the most sophisticated means to determine the effective electron antineutrino mass; it is expressed as $m^2_\beta = \sum_i m_i^2 |U_{ei}|^2$. The second method relies on the search for $0\nu \beta \beta$ processes whose decay amplitude is proportional to the effective Majorana neutrino mass defined as $|m_{\beta\beta}| = |\sum_i U_{ei}^2 m_i|$. These two parameters can be represented as functions of the model parameters by substituting the neutrino masses with their expressions from Eq. \ref{ev} in the following relations
\begin{equation}
    m^2_\beta = \frac{1}{3} \left(2 m^2_1 \cos^2\theta + m^2_2 + 2 m^2_3 \sin^2\theta\right), \quad |m_{\beta\beta}| = \frac{1}{3} \left|2 m_1 \cos^2\theta + m_2 e^{i\alpha} + 2 m_3 \sin^2\theta e^{i(\beta - 2\sigma)}\right|.
    \label{mbb}
\end{equation}
It is important to note that the $0\nu \beta \beta$ decay processes serve as a means to probe the Majorana nature of neutrinos. Consequently, the expression for $|m_{\beta \beta}|$ in Eq. \ref{mbb} incorporates two additional Majorana phases, namely $\alpha$ and $\beta$. In Fig. \ref{pheno}, the middle panel displays the correlation between $m_\beta$ and $m_1$, and the right panel shows the correlation between $m_{\beta\beta}$ and $m_1$. These scatter plots emerge from varying the oscillation parameters within their $3\sigma$ range, while the Majorana phases are varied in the range of $[0,2\pi]$. The predicted values for these observables are  given by
\begin{equation}
    0.009 \lesssim m_\beta[eV] \lesssim 0.031, \quad 0.000049 \lesssim |m_{\beta\beta}|[eV] \lesssim 0.02963
\end{equation}
The values of $m_\beta$ are considerably smaller than the KATRIN sensitivity ($\sim 0.2~eV$) \cite{kat}, making them challenging to test in the near future. Nevertheless, the upper limit of $m_\beta$ is close to the expected sensitivity of the Project 8 collaboration $\sim 0.04~eV$ (depicted by the blue dashed line in the middle panel of Fig. \ref{pheno}) \cite{P8}. Consequently, future experiments with improved sensitivities could potentially reach this upper bound. For $\left\vert m_{\beta \beta }\right\vert $, the dashed lines in the right panel of Fig. \ref{pheno} show that our model values for are currently lower than the sensitivities of some ongoing experiments. On the other hand, upcoming experiments like nEXO \cite{nex} and GERDA Phase II \cite{ger} are expected to be able to test our predicted values of $\left\vert m_{\beta \beta }\right\vert $.
\subsection{Solution to the DW problem}
As mentioned earlier, DWs are topological defects with surface-like characteristics that arise during the early stages of the Universe when a discrete symmetry undergoes spontaneous breaking \cite{zel,kib}. In this work, we focused on two $A_4$ breaking patterns resulting in the formation of DWs in the charged lepton and neutrino sectors. Typically, the presence of DWs becomes problematic when the scale of the SSB is lower than the inflationary scale $\sim (10^{14}-10^{16})~GeV$ \cite{inf0,inf1,inf2,inf3}. Let us start by estimating the scale of symmetry breaking that takes place in the charged lepton sector, precisely in the framework of $A_4 \rightarrow Z^T_3$. Recall that this breaking pattern occurs when the flavon triplet $\Phi$ attains a VEV along the direction $\left\langle \Phi \right\rangle = \upsilon_\Phi (1,0,0)^T$, where the broken part is given by the Klein four group $K_4$. Consequently, there exist four degenerate vacua \{$\phi_1,\phi_2,\phi_3,\phi_4$\} associated with this breaking pattern; see Eq. \ref{phi}. Furthermore, this breaking pattern yielded the charged lepton masses in Eq. \ref{ml+}. By examining, for instance, the expression of the tau lepton mass ($m_\tau = y_\tau \upsilon_d \upsilon_\Phi / \Lambda$), we can derive an approximation for the ratio $\upsilon_\Phi / \Lambda$.
Employing the value of the tau lepton mass from the PDG, given as $m_\tau = 1776.85~MeV$ \cite{PDG}, and under the reasonable assumption that $y_\tau \upsilon_d \lesssim 246~GeV$, we establish a lower limit for this ratio, which is expressed as $\upsilon_\Phi / \Lambda > 0.007$. Assuming that $\Lambda$ is approximately the SUSY-GUT scale $M_{GUT} \sim 2 \times 10^{16}~GeV$, we obtain a lower bound for the flavon VEV given as $\Phi > 1.4 \times 10^{14}~GeV$. Consequently, in the charged lepton sector, the discrete $A_4$ symmetry breaks down during the inflationary scale. In this case, any produced DWs are inflated away, leaving only the $Z^T_3$ invariant ground state $\phi_1 = \left\langle \Phi \right \rangle$.

Next, we analyze the DWs in the neutrino sector and estimate the scale of the symmetry breaking pattern $A_4 \rightarrow Z^S_2$ in which the broken part is given by $Z^{TST^2}_2 \rtimes Z^T_3$. The Majorana mass terms in the superpotential \ref{wy} are at the renormalizable level. We expect a hierarchy among the VEVs of the NMSSM singlet and the flavon superfields with $\upsilon_S < \upsilon_\chi \lesssim \upsilon_\Omega$. This hierarchy makes sense because $\upsilon_S$ is known to acquire a VEV around the scale of $M_{SUSY}$. Additionally, it plays a role in introducing a slight deviation from the TBM scheme, implying that its VEV should be smaller than those of the flavon fields. Consequently, the VEVs of flavons $\chi$ and $\Omega$ can be reasonably placed in a range above $M_{SUSY}$, say a range from $10^7$ to $10^{10}~GeV$. This range ensures that the right-handed neutrino masses ($M_i$) in Eq. \ref{maj} align with the type I seesaw mechanism. Achieving this involves selecting the appropriate values for the free parameters $Y_0$, $\lambda_\chi$, $\lambda_\Omega$, and $\beta$ in $\upsilon_u=\upsilon \sin\beta$ with $\upsilon = 246~GeV$. As a result, these VEVs are smaller than the inflationary scale, leading to the inevitable creation of stable DWs in the neutrino sector. These walls are inconsistent with standard cosmology and must be avoided \cite{wmap}. This is known as the DW problem. \newline
To address this issue, we follow the well-known solution suggested by Zel'dovich et al. \cite{zel}, which relies on making the walls unstable by assuming that the concerned discrete symmetry is only approximate. Specifically, introducing terms that break explicitly the discrete symmetry will create a small energy  difference between the vacua. This can be achieved by introducing higher dimensional operators $\frac{1}{M^n_{Pl}}O_{n+3}$ suppressed by powers of the Planck scale $M_{Pl}$ leading to favor one of the vacua over the others, and consequently the false vacua disappear before the walls take over as the main energy source in the Universe, see Refs. \cite{A5,B3,A11} for more details and Ref. \cite{NMSSM} for the case of NMSSM with $Z_3$ discrete symmetry. 
\newline
In our case, the leading order operators that break explicitly the $Z^{TST^2}_2 \rtimes Z^T_3  \subset A_4$ are of order five. Many operators fall into this category, and all the possible terms can be chosen as given by a perturbation $\delta W_{scal}$ applied to the trilinear couplings that are present in the $A_4 \rtimes Z_3$ invariant superpotential\footnote{It is crucial to emphasize that this scalar superpotential is restricted to the Higgs doublets $H_{u,d}$, the gauge singlet $\mathcal{S}$, and the flavons $\chi$, $\Omega$, and $\Phi$.} of the model given by
\begin{equation}
    W_{scal} = \mu^2 \Phi^2 + \lambda_1 \mathcal{S} H_u H_d + \lambda_2 \mathcal{S}^3 + \lambda_3 \Omega^3 + \lambda_4 \Phi^3 + \lambda_5 \chi^3 + \lambda_6 \mathcal{S} \Omega^2 + \lambda_7 \chi \Omega^2.
\end{equation}
The most appropriate five dimensional operator that can break the full discrete symmetry group $A_4 \times Z_3$ is expressed as
\begin{equation}
    W_{NR} = \frac{\lambda^{\prime}_3}{M_{Pl}^2} (\Omega^5)|_{(1,\omega)}
    \label{nr}
\end{equation}
which transforms as a nontrivial $A_4$ singlet. Since $\Omega$ transforms as $\sim \mathbf{1}_{Q^2}$ under the extra $Z_3$ symmetry, this operator also induces the breaking of the latter\footnote{The other higher dimensional operators of order five that can break explicitly the full flavor group are shown in Appendix C of Ref, \cite{A4}}.
\begin{figure}[thb]
\centering
\includegraphics[width=0.18\linewidth]{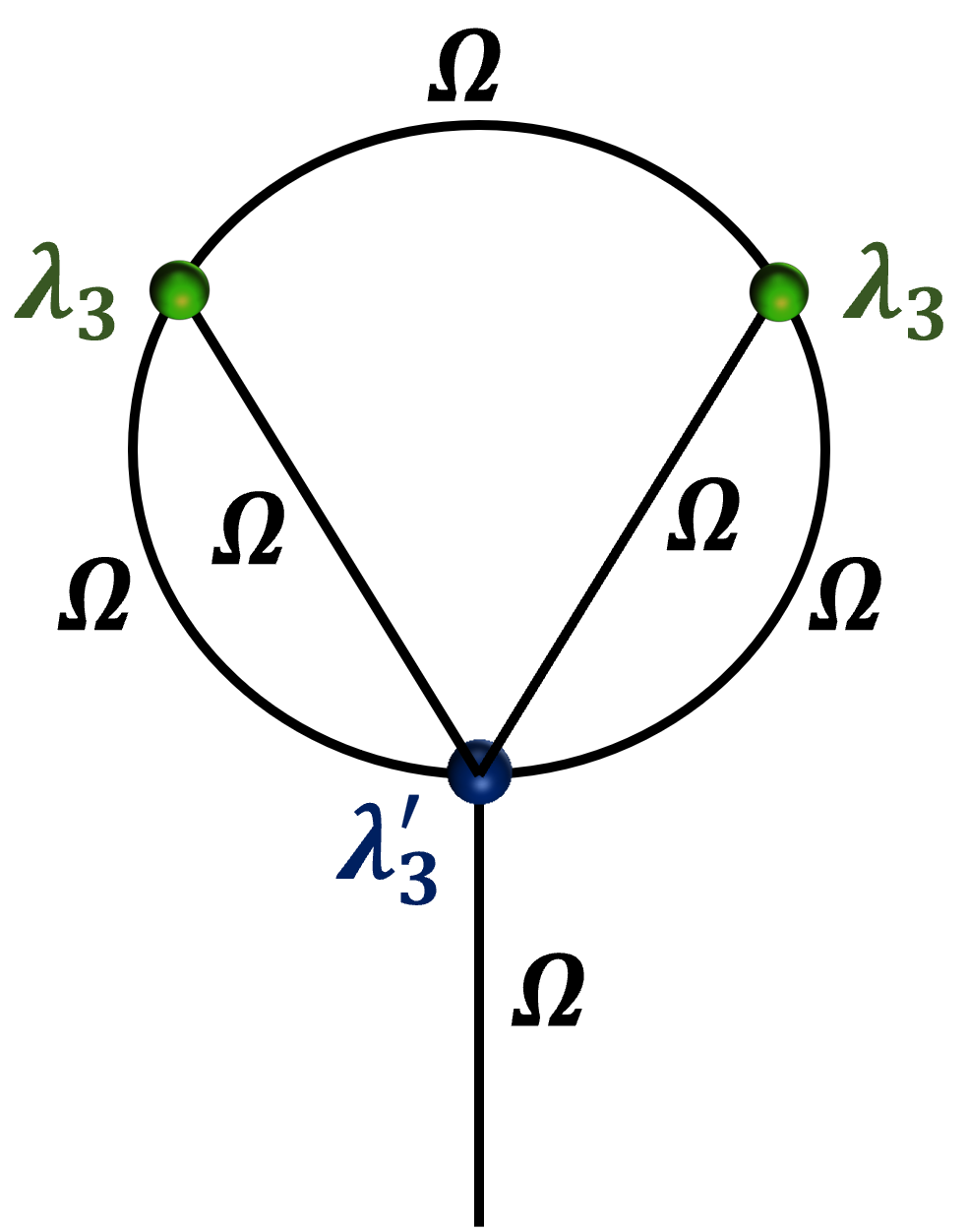}
\caption{The tadpole diagram for the higher dimensional operator in
Eq. \ref{nr}.}
\label{tdi}
\end{figure}
At the quantum level, the operator in Eq. \ref{nr} is represented by the Feynman diagram in Fig. \ref{tdi}. Thus, By using the Feynman rules of supergraphs \cite{sg1,sg2,sg3}, we obtain the following contribution to the effective scalar potential\footnote{Details calculations on how to arrive at this effective potential are provided in appendix C of \cite{A4}. See also Refs. \cite{sg1,sg2,sg3}}

Tthe operator described by Eq. \ref{nr} can be represented by the Feynman diagram illustrated in Fig. \ref{tdi}. Employing the Feynman rules of supergraphs \cite{sg1,sg2,sg3}, we derive the corresponding contribution to the effective scalar potential\footnote{Detailed calculations leading to the formulation of this effective potential can be found in the appendix C of \cite{A4}, see also Refs. \cite{sg1,sg2,sg3}.}
\begin{equation}
    \delta V_{eff} = M^3_W(\eta \phi_\Omega + \Bar{\eta} \Bar{\phi_\Omega}) + M^2_W(\eta F_\Omega + \Bar{\eta} \Bar{F_\Omega} )
\end{equation}
where $\eta = \frac{\lambda^2_3 \lambda^{\prime}_3}{(16 \pi^2)^3}$ and $M_W$ is the scale of the electroweak theory, while $\phi_\Omega$ and $F_\Omega$ are the scalar component and the F-term of the flavon superfield $\Omega$. Accordingly, the higher dimensional operator in Eq. \ref{nr} induces a linear term in the soft SUSY scalar potential expressed as $V_{soft} \supset \eta M^3_W \phi_\Omega + h.c$. Due to the nontrivial transformation of the triplet $\phi_\Omega$ under the $A_4 \times Z_3$ flavor symmetry, the term in $V_{soft}$ breaks explicitly the symmetry into the $Z_2$ group. This contribution induces an energy difference among the six degenerate vacua in Eq. \ref{phipr}. The energetically dominant true vacuum, say $\phi^{\prime}_1$, becomes the domainat state causing the remaining false vacua to disappear before the walls dominate the energy density of the Universe.
\section{Summary and conclusion}
In this work, we have studied two primary aspects of SSB of non-Abelian discrete symmetries. In the first part, we investigate neutrino phenomenology within a flavored NMSSM based on the flavor symmetry $A_4 \times Z_3$. This involves extending the usual NMSSM by incorporating three right-handed neutrino and two flavon superfields. Each superfield is assigned representations under the $A_4 \times Z_3$ group to engineer appropriate neutrino masses and mixing. Notably, we achieve this in a minimal manner in the sense that we employ only two flavon fields and the NMSSM singlet $\mathcal{S}$ responsible for the deviation from the well-known TBM pattern. Then, we delved into the phenomenological implications of the neutrino sector, particularly in the NH scheme. By using the $3\sigma$ allowed ranges of the oscillation parameters, we showed through scatter plots the permissible ranges for various observables, such as $\sum_i m_i$, $m_{\beta\beta}$, and $m_\beta$.
In the second part, our investigation focused on the breaking patterns of the $A_4$ flavor group and the formation of DWs as a result of its SSB. We showed for the breaking patterns $A_4 \rightarrow Z_3$ (charged lepton sector) and $A_4 \rightarrow Z_3$ (neutrino sector) the explicit expressions of the degenerate vacua. Then, after representing these vacua as vertices in flavon space, we depicted the DW networks by connecting them. The resultant configuration gives rise to Platonic solids, where vertices representing degenerate vacua are linked by edges symbolizing DWs. By analyzing the breaking scale of the flavor symmetry, we found that the DWs formed in the neutrino sector conflict with cosmological observations. To address this issue, we used the well-known solution of breaking the flavor symmetry explicitly employing higher dimensional Planck suppressed operators.

\section*{Acknowledgement}
This work is supported by the United Arab Emirates University (UAEU) under UPAR Grant No. 12S093.

\appendix
\section{The $A_4$ group}
In this appendix, we present some pertinent properties associated with the $A_4$ flavor group, which prove to be indispensable in the study of fermion masses and mixing as well as the formation of the DWs. The non-Abelian group $A_4$ is generated by two noncommuting generators $S$ and $T$ satisfying $S^2 = T^3 = (ST)^3 = I$. In the Altarelli-Feruglio basis where the generator $T$ is diagonal, the three-dimensional representation matrices of $S$ and $T$ are given by \cite{AF}
\begin{equation}
T = \begin{pmatrix}
    1 & 0 & 0\\
    0 & \omega^2 & 0\\
    0 & 0 & \omega
    \end{pmatrix}   , \quad
S = \begin{pmatrix}
    -1 & 2 & 2\\
    2 & -1 & 2\\
    2 & 2 & -1
    \end{pmatrix}, 
    \label{basis}
\end{equation}
where $\omega = e^{2\pi i/3}$. The group $A_4$ has four irreducible representations: one triplet $\mathbf{3}_{(-1,0)}$ and three different singlets $\mathbf{1}_{(1,1)}$, $\mathbf{1}_{(1,\omega)}$ and $\mathbf{1}_{(1,\omega^2)}$. The indices in these representations correspond to the characters of the generators of $A_4$ where the first entry represents the characters of $S$ generator, while the second entry represents the characters of the $T$ generator. For the one-dimensional representations, these generators are expressed as
\begin{equation}
    \mathbf{1}_{(1,1)}: S=1, \quad T=1; \quad \mathbf{1}_{(1,\omega)}: S=1, \quad T=\omega^2; \quad \mathbf{1}_{(1,\omega^2)}: S=1, \quad T=\omega.
\end{equation}

The order of the $A_4$ group is 12 and it possesses four distinct conjugacy classes given by
\begin{equation}
   C_1 = \{e\}, \quad C_2 = \{S, TST^{-1}, T^{-1}ST \},\quad 
   C_3 = \{T, TS, ST, STS \},\quad C_4 = \{T^2, ST^2, T^2 S, ST^2S \},
\end{equation}
thereby providing the representation of the 12 elements within $A_4$. Besides the elements of $A_4$, a crucial aspect for investigating the formation of DWs is identifying all the subgroups within the $A_4$ symmetry. In addition to the trivial subgroup, the identity $I$, and the whole group itself, $A_4$ has three cyclic $Z_2$ subgroups, four cyclic $Z_3$ subgroups and one Klein-four group $K_4$. In terms of the elements of $A_4$, these subgroups are given by
\begin{equation}
    \begin{aligned}
        Z_2^S &= \{I, S \}, \quad Z_2^{TST^{-1}} = \{I, TST^{-1} \}, \quad Z_2^{T^2ST} = \{I, T^2ST \}, \\
        Z_3^T &= \{I, T, T^2 \}, \quad Z_3^{ST} = \{I, ST, T^2S \}, \quad Z_3^{TS} = \{I, TS, ST^2 \}, \quad Z_3^{STS} = \{I, STS, ST^2S \}, \\
        K_4^{S, TST^2} &= \{I, S, TST^2, T^2ST \}
    \end{aligned}
\end{equation}

Let us now outline the rules governing the tensor product of irreducible representations within the $A_4$ group. These representations are denoted by their basis characters and given as follows
\begin{equation}
    \mathbf{1}_{(1,\omega^i)} \times \mathbf{1}_{(1,\omega^j)} = \mathbf{1}_{(1,\omega^{i+j})},\quad \mathbf{3}_{(-1,0)} \times \mathbf{1}_{(1,\omega^i)} = \mathbf{3}_{(-1,0)},\quad \mathbf{3}_{(-1,0)} \times \mathbf{3}_{(-1,0)} = \mathbf{1}_{(1,1)} + \mathbf{1}_{(1,\omega)} + \mathbf{1}_{(1,\omega^2)} + \mathbf{3}^S_{(-1,0)} + \mathbf{3}^A_{(-1,0)},
\end{equation}
where $i,j = 0,1,2$. The tensor product of two $A_4$ triplets, denoted as $(a_1,a_2,a_3)^T$ and $(b_1,b_2,b_3)^T$, is expressed explicitly as
\begin{eqnarray}
    \begin{pmatrix} a_1 \\ a_2 \\ a_3 \end{pmatrix} \otimes \begin{pmatrix} b_1 \\ b_2 \\ b_3 \end{pmatrix} &=& (a_1 b_1 + a_2 b_3 + a_3 b_2)_{(1,1)} + (a_3 b_3 + a_2 b_1 + a_1 b_2)_{(1,\omega)} + (a_2 b_2 + a_1 b_3 + a_3 b_1)_{(1,\omega^2)} \nonumber \\
    &+& \frac{1}{3} \begin{pmatrix} 2 a_1 b_1 - a_2 b_3 - a_3 b_2 \\ 2 a_3 b_3 - a_2 b_1 - a_1 b_2 \\ 2 a_2 b_2 - a_1 b_3 - a_3 b_1 \end{pmatrix}_{ \mathbf{3}^S_{(-1,0)}} + \frac{1}{2} \begin{pmatrix} a_2 b_3 - a_3 b_2 \\ a_1 b_2 - a_2 b_1 \\ a_3 b_1 - a_1 b_3 \end{pmatrix}_{\mathbf{3}^A_{(-1,0)}}.
\end{eqnarray}

\bibliographystyle{JHEP}
\bibliography{bibliography.bib}

\end{document}